\documentclass[prd,aps,preprintnumbers,floats,floatfix,superscriptaddress,preprintnumbers,
showpacs,eqsecnum,nofootinbib,twocolumn]{revtex4-1}
\usepackage[utf8]{inputenc}
\usepackage{latexsym,array,theorem,mathrsfs,bm,float}
\usepackage{psfrag}
\usepackage{amsfonts,amsmath,amssymb,latexsym,array,afterpage,
theorem,mathrsfs,bm,float,epsfig,color,graphicx,tabularx,here,multirow}

\newcommand{\nn}{\nonumber \\}
\newcommand{\bea}{\begin{eqnarray}}
\newcommand{\ena}{\end{eqnarray}}
\newcommand{\beann}{\begin{eqnarray*}}
\newcommand{\enann}{\end{eqnarray*}}

\begin{document}
\baselineskip=12pt

%<<<<<<<<<<<<< TITLE >>>>>>>>>>>>>>>%
%%
\title{Vainshtein mechanism in massive gravity nonlinear sigma models
%$\Lambda_2$ decoupling limit
}
%%
%<<<<<<<<<<<<< AUTHOR >>>>>>>>>>>>>>>%
%%
\author{Katsuki \sc{Aoki}}
\email{katsuki-a12@gravity.phys.waseda.ac.jp}
\affiliation{
Department of Physics, Waseda University,
Shinjuku, Tokyo 169-8555, Japan
}

\author{Shuntaro \sc{Mizuno}}
\email{shuntaro.mizuno@aoni.waseda.jp}
\affiliation{Waseda Institute for Advanced Study, Waseda University, Shinjuku, Tokyo 169-8050, Japan}

%<<<<<<<<<<<<< DATE >>>>>>>>>>>>>>>%
\date{\today}

%======================================%
%<<<<<<<<<<<<< ABSTRACT >>>>>>>>>>>>>>>%
%======================================%
\begin{abstract}
We study the stability of the Vainshtein screening solution of the massive/bi-gravity based on the massive nonlinear sigma model as
the effective action inside the Vainshtein radius. The effective action is obtained by taking the $\Lambda_2$ decoupling limit around a curved spacetime. First we derive a general consequence that any Ricci flat
Vainshtein screening solution is 
unstable when we take into account the excitation of the scalar graviton only.
This instability suggests that the nonlinear excitation of the scalar graviton is not sufficient to obtain a successful Vainshtein screening in massive/bi-gravity.
Then  to see the role of the excitation of the vector graviton, we study perturbations around
the static and spherically symmetric solution obtained in bigravity explicitly.
As a result, we find that linear excitations of the vector graviton cannot be helpful and the solution still suffers from a ghost and/or a gradient instability for any parameters of the theory for this background.
\end{abstract}

%<<<<<<<<<<<<< PACS NUMBER >>>>>>>>>>>>>>>%
%\pacs{04.60.Cf, 04.50.Gh, 04.50.-h, 11.25.-w }

% 04.40.Dg	Relativistic stars: structure, stability, and oscillations (see also 97.60.-s Late stages of stellar evolution)
% 04.50.Gh : higher-dimensional Black holes
% 04.50.-h : Higher-dimensional gravity and other theories of gravity
% 04.60.Cf : gravitational aspects of String theory
% 11.25.-w : Strings and branes
% 04.50.Kd : Modified theories of gravity
% 98.80.-k : Cosmology
% 95.36.+x : Dark energy

\maketitle

\section{Introduction}
The current acceleration of the Universe is one of the biggest problems in the modern cosmology. It has been proposed to explain this acceleration with a modification of the theory of gravity from general relativity (GR) at the infrared regime  (see \cite{Clifton:2011jh,Joyce:2014kja,Koyama:2015vza} for reviews). However the modification of the gravity is strongly constrained by Solar System tests of the gravity which agree with the predictions of GR. Hence the effect of the modification of gravity must be screened at Solar System. One natural theory with such a screening mechanism is the massive gravity with a tiny graviton mass. In short scales the massive graviton may behave as a massless graviton, thus one expects the prediction of GR can be recovered. However the linear massive gravity \cite{Fierz:1939ix} cannot be restored to the linearized GR because of the non-vanishing fifth force mediated by the scalar graviton \cite{vanDam:1970vg,Zakharov:1970cc}. Vainshtein then proposed that the linear approximation is no longer valid inside the Vainshtein radius and the fifth force could be screened by nonlinear interactions, called the Vainshtein mechanism \cite{Vainshtein:1972sx}. 

The Vainshtein mechanism has an important role not only in the massive gravity but also in some classes of
the scalar-tensor theories which have non-linear interactions \cite{Nicolis:2008in,Deffayet:2009wt,Deffayet:2009mn,Deffayet:2011gz,Kobayashi:2011nu,Horndeski:1974wa,Gleyzes:2014dya,Gleyzes:2014qga}. However because of their nonlinear interactions, a general analysis of the Vainshtein mechanism is quite complicated. Hence to discuss the Vainshtein mechanism, it should be useful to construct an effective theory from an original theory and discuss the Vainshtein mechanism based on the effective theory \cite{Kimura:2011dc,Koyama:2013paa,Kobayashi:2014ida,Saito:2015fza}. 

In this paper, we discuss an effective theory for the Vainshtein mechanism from the nonlinear massive gravity \cite{deRham:2010ik,deRham:2010kj} and the bigravity \cite{Hassan:2011zd} (see \cite{Hinterbichler:2011tt,Babichev:2013usa,deRham:2014zqa,Schmidt-May:2015vnx} for reviews). The Vainshtein mechanism in massive/bi-gravity has been discussed in \cite{Babichev:2009us,Babichev:2009jt,Babichev:2010jd,Koyama:2011xz,Nieuwenhuizen:2011sq,Koyama:2011yg,Chkareuli:2011te,Gruzinov:2011mm,Babichev:2011iz,Comelli:2011wq,Berezhiani:2011mt,Sjors:2011iv,Volkov:2012wp,Sbisa:2012zk,Volkov:2013roa,Babichev:2013pfa,Kaloper:2014vqa,Renaux-Petel:2014pja,Enander:2015kda,Aoki:2015xqa,Aoki:2016eov}. It is known that the nonlinear massive gravity can be reduced into the scalar-tensor theory with Galileon interactions by taking the $\Lambda_3$ decoupling limit when the vector graviton is not excited in which there still exist scalar-tensor interactions  \cite{ArkaniHamed:2002sp,deRham:2010ik} (see also \cite{Ondo:2013wka,Fasiello:2013woa}). On the other hand, one can take more direct decoupling limit called $\Lambda_2$ decoupling limit \cite{deRham:2015ijs,deRham:2016plk} in which the tensor fluctuations are decoupled from the scalar and vector gravitons and the effective action for the scalar and vector gravitons is given by the massive gravity nonlinear sigma model. The papers \cite{deRham:2015ijs,deRham:2016plk} discussed the $\Lambda_2$ decoupling limit  around the Minkowski spacetime.  Contrary to this, in the present paper, we discuss the $\Lambda_2$ decoupling limit around a curved spacetime and obtain an effective theory inside the Vainshtein radius. Indeed the solution obtained by the effective theory gives an approximate solution inside the Vainshtein radius in the bigravity theory \cite{Aoki:2016eov}.  Then we study the stability of the Vainshtein screening solution
%mechanism 
based on the effective theory.

The paper is organized as follows. We derive the effective theory for the Vainshtein mechanism from  massive/bi-gravity in Section \ref{sec_ET_Vainshtein}.  In Section \ref{sec_scalar_instability} we then study the dynamics of the scalar graviton around general backgrounds and find a Ricci flat spacetime generally suffers from a ghost and/or a gradient instability. The instability is found when we ignore the vector graviton, however perturbations of scalar and vector gravitons are coupled to each other, in general. Hence to complete the stability analysis of
the solution,  
we should include perturbations of the vector graviton, which is studied in Section \ref{sec_instability_SSS}. We explicitly show that the static and spherically symmetric solution is unstable. We give a summary and some discussions in Section \ref{summary}.

\section{Effective action inside Vainshtein radius}
\label{sec_ET_Vainshtein}
 In this section, we show that massive gravity nonlinear sigma model gives an effective theory of the vector and scalar gravitons
inside the Vainshtein radius for general massive/bi-gravity as long as we have the Vainshtein screening solutions.
Let us start with the action  \cite{Hassan:2011zd}  
given by
\begin{align}
\!\!\!\!\!\!\!\!\!\!  S &=\frac{1}{2 \kappa _g^2} \int d^4x \sqrt{-g}R(g)+ \frac{1}{2 \kappa _f^2}
 \int d^4x \sqrt{-f} \mathcal{R}(f) \nonumber \\
&-
\frac{m^2}{ \kappa ^2} \int d^4x \sqrt{-g} \sum_{n=2}^{4} c_n \mathscr{U}_n(\gamma) 
+S^{[\rm m]}\,,
\label{action}
\end{align}
where $g_{\mu\nu}$ and $f_{\mu\nu}$ are two dynamical metrics, and
$R(g)$ and $\mathcal{R}(f)$ are their Ricci scalars. The interactions between two metrics are given by
\begin{align}
\mathscr{U}_2(\mathcal{K})&=-\frac{1}{4}\epsilon_{\mu\nu\rho\sigma} 
\epsilon^{\alpha\beta\rho\sigma}
{\mathcal{K}^{\mu}}_{\alpha}{\mathcal{K} ^{\nu}}_{\beta}\,, \nn
\mathscr{U}_3(\mathcal{K})&=-\frac{1}{3!}\epsilon_{\mu\nu\rho\sigma} 
\epsilon^{\alpha\beta\gamma\sigma}
{\mathcal{K} ^{\mu}}_{\alpha}{\mathcal{K} ^{\nu}}_{\beta}{\mathcal{K} ^{\rho}}_{\gamma}\,, 
\\
\mathscr{U}_4(\mathcal{K})&=-\frac{1}{4!}\epsilon_{\mu\nu\rho\sigma} 
\epsilon^{\alpha\beta\gamma\delta}
{\mathcal{K} ^{\mu}}_{\alpha}{\mathcal{K} ^{\nu}}_{\beta}{\mathcal{K} ^{\rho}}_{\gamma}
{\mathcal{K}^{\sigma}}_{\delta}\,,
\nonumber
\end{align}
with $\mathcal{K}^{\mu}{}_{\nu}=\delta^{\mu}{}_{\nu}-\gamma{}^{\mu}{}_{\nu}$ and ${\gamma^{\mu}}_{\nu}$ is 
defined by 
\begin{equation}
{\gamma^{\mu}}_{\rho}{\gamma^{\rho}}_{\nu}
=g^{\mu\rho}f_{\rho\nu}
\,.
\label{gamma2_metric}
\end{equation}
The parameters  $\kappa_g^2=8\pi G$ and $\kappa_f^2=8\pi \mathcal{G}$ are 
the corresponding gravitational constants, 
while $\kappa$ is defined by $\kappa^2=\kappa_g^2+\kappa_f^2$. 

The matter action is assumed such that a matter field can couple with either $g_{\mu\nu}$ or $f_{\mu\nu}$:
\begin{align}
S^{[\rm m]}=S^{[\rm m]}_g(\psi_g,g)+S_f^{[\rm m]}(\psi_f,f)\,.
\end{align}
Notice that our set-up of the bigravity is so general that it includes ghost-free massive gravity \cite{deRham:2010ik,deRham:2010kj} as a special case.
It is obtained by fixing $f$-spacetime as the Minkowski one with the limit $\kappa_f \rightarrow 0$ \cite{Baccetti:2012bk}.

In what follows we set $c_2=-1$, which guarantees the parameter $m$ corresponds to the graviton mass propagating on the Minkowski vacuum\footnote{In general, the ghost-free interactions include a constant term $\mathscr{U}_0$ and a tadpole term $\mathscr{U}_1(\mathcal{K})$. 
 Although we drop them in this paper, just for simplicity, including these terms does not change our main conclusion.
}.
 We also introduce
four St\"ueckelberg fields $\phi^a(x)$, with which the metric $f_{\mu\nu}$ can be written by
\begin{align}
f_{\mu\nu}(x)=\frac{\partial \phi^a}{\partial x^{\mu}} \frac{\partial \phi^b}{\partial x^{\nu}} f_{ab}(\phi^a(x))
\,,
\label{fmunu}
\end{align}
 to see the dynamics of the vector and scalar gravitons in a clear way.

 It was shown in Refs.~\cite{deRham:2015ijs,deRham:2016plk}  that in the case of ghost-free massive gravity, about non-trivial vacua
\begin{align}
g_{\mu\nu} = \eta_{\mu\nu} + \mathcal{O} (m^2)\,,\;\;\;\;\phi^a = \bar{\phi}^a (x) \neq x^a\,, 
\end{align} 
and in the $\Lambda_2$ decoupling limit, given by
\begin{align}
m,\kappa_g,\kappa_f\rightarrow0\,,\quad
\Lambda_2 \equiv \sqrt{m/\kappa_g}:{\rm finite}\,,
\label{Lambda2_decoupling}
\end{align}
an interesting effective theory for $\phi^a$, so-called the massive gravity nonlinear sigma model described by the following action
\begin{align}
S^{(0)}_{\rm MG-NLS}=-\Lambda_2^4 \int d^4x \sqrt{- \eta }\sum_{n=2}^{4} c_n \mathscr{U}_n(\gamma_{\rm NLS})
\,,
\label{NLSM_action_flat}
\end{align}
\begin{align}
\gamma_{\rm NLS}{}^{\mu}{}_{\rho}\gamma_{\rm NLS}{}^{\rho}{}_{\nu}=
\eta^{\mu\rho}(x)
\frac{\partial \phi^a}{\partial x^{\rho}} \frac{\partial \phi^b}{\partial x^{\nu}} 
\eta_{ab}(\phi)\,.
\end{align}
is obtained.

One interesting property with this massive gravity nonlinear sigma model is that its strong coupling scale is given by $\Lambda_2$.
This is higher than $\Lambda_3 \equiv (m^2/\kappa_g)^{1/3}$, considered as the highest possible strong coupling scale in ghost-free massive gravity,
coming from the analysis around the trivial vacuum $g_{\mu\nu} = \eta_{\mu\nu}$, $\phi^a = x^a$.
Another interesting property is that the vector and scalar modes of graviton encoded in $\phi^a$ decouple with matter fields even in the linear regime,
which does not give vDVZ discontinuity and the Vainshtein mechanism is implemented automatically.

This suggests that as long as the Vainshtein screening works,
even if we start with more general set-up described by Eq.~(\ref{action}), that is, not limiting $g_{\mu\nu}$ and $f_{ab}$ to flat,
not neglecting the  $f$-matter fields, we can expect that the massive gravity nonlinear sigma model is obtained as an effective theory
inside the Vainshtein radius.
Actually, if the Vainshtein mechanism is working, the metrics can be expressed by
\begin{align}
g_{\mu\nu}&=g_{\mu\nu}^{\rm GR}+\kappa_g \delta g_{\mu\nu}
\,,
\\
f_{ab}&=f_{ab}^{\rm GR}+\kappa_f \delta f_{ab}
\,,
\end{align}
where $\delta g_{\mu\nu}$ and $\delta f_{ab}$ should be treated as perturbations. 
The metrics $g_{\mu\nu}^{\rm GR}$ and $f_{ab}^{\rm GR}$ are assumed to be solutions in GR with the matter actions $S_g^{[\rm m]}$ and $S_f^{[\rm m]}$, respectively,
in order to remove the tadpole terms for $\delta g_{\mu\nu}$ and $\delta f_{ab}$ in the $\Lambda_2$ decoupling limit\footnote{In the case of massive gravity, the $f$-metric can be a generic metric rather than a GR solution since the Einstein-Hilbert action for $f_{\mu\nu}$ is absent.}.
%$g_{\mu\nu}^{\rm GR}$ and $f_{ab}^{\rm GR}$ are solutions in GR with the matter actions $S_g^{[\rm m]}$ and $S_f^{[\rm m]}$, respectively,
%and $\delta g_{\mu\nu}$ and $\delta f_{ab}$ should be treated as perturbations. 
Here, 
$g_{\mu\nu}^{\rm GR}(x)$ and $f_{ab}^{\rm GR}(\phi)$ are
determined
as functions of $x^{\mu}$ and $\phi^a$, respectively,
from which  we can regard that
$g_{\mu\nu}^{\rm GR}$ and $f_{ab}^{\rm GR}$ act as external forces.
Then  with the undetermined variables  $\phi^a,\delta g_{\mu\nu}, \delta f_{ab}$, the action can
be expanded as
\begin{align}
S&=S_{\rm GR}(\delta g)+S_{\rm GR}(\delta f)
\nn
&+S_{\rm MG-NLS}(\phi^a;g^{\rm GR},f^{\rm GR})
+\Lambda_2^4 \mathcal{O}(\kappa_g \delta g,\kappa_f \delta f)
\,.
\end{align} 
where $S_{\rm GR}$ are the perturbed actions for the metric perturbations which are same as those in GR.
%and
$S_{\rm MG-NLS}$ is the action of the massive gravity nonlinear sigma model given by
\begin{align}
S_{\rm MG-NLS}=-\Lambda_2^4 \int d^4x \sqrt{-g_{\rm GR} }\sum_{n=2}^{4} c_n \mathscr{U}_n(\gamma_{\rm NLS})
\,,
\label{NLSM_action}
\end{align} 
which generalizes Eq.~(\ref{NLSM_action}) with the following replacements:
\begin{align}
\kappa_g \to \kappa\,,\;\;\;\;\eta_{\mu\nu} \to g_{\mu\nu}^{\rm GR}\,,\;\;\;\; \eta_{ab} \to f_{ab}^{\rm GR}\,.
\label{NLSM_action_generalizations}
\end{align}

One may worry that the tadpole terms of the metric perturbations give the backreaction from the  St\"ueckelberg fields to the spacetimes.
Since the background spacetimes are given by the solutions in GR, 
they appear only through 
the interaction terms between the St\"ueckelberg fields and the metric perturbations of order
$\Lambda_2^4\mathcal{O}(\kappa_g \delta g,\kappa_f \delta f)$.
However, by taking the $\Lambda_2$ decoupling limit  given by Eq.~(\ref{Lambda2_decoupling}),
the contributions form the tadpole terms are negligible and then the St\"ueckelberg fields and the metric perturbations are decoupled.
Then, in this limit the St\"ueckelberg fields are simply determined by the massive gravity nonlinear sigma model \eqref{NLSM_action} and the spacetimes are completely same as those in GR.
Therefore the massive gravity nonlinear sigma model with curved metrics is 
the effective action of the St\"ueckelberg fields 
as long as the Vainshtein mechanism works and we have the same solutions as in GR. 

Indeed, the Vainshtein screening solutions can be obtained by this effective action with curved metrics.
The Vainshtein mechanism for the static and spherically symmetric spacetime is found with the interior and the exterior Schwarzschild metrics \cite{Aoki:2016eov} (see also \cite{Babichev:2013pfa,Enander:2015kda}) and the cosmological Vainshtein mechanism is found with the Friedmann-Lema$\hat{\i}$tre-Robertson-Walker metric \cite{Aoki:2015xqa}. Although we denoted the procedure of the $\Lambda_2$ decoupling limit just as the massless limit in these papers, the limits used in \cite{Aoki:2015xqa,Aoki:2016eov} are equivalent to the $\Lambda_2$ decoupling limit shown above.

For simplicity, we do not introduce the $f$-matter fields  in this paper thus we can assume $f_{ab}^{\rm GR}$ is the Minkowski spacetime.
In this case,  the St\"ueckelberg field can be split as $\phi^a=\delta^{a}_\mu(x^\mu+\pi^\mu)$,
then $f_{\mu\nu}$ is expressed by
\begin{align}
f_{\mu\nu}=\eta_{\mu\nu}+2\partial_{(\mu}\pi_{\nu)}
+\partial_{\mu}\pi_{\alpha}\partial_{\nu}\pi^{\alpha}\,,
\end{align}
where $\pi_{\mu}$ is a vector field on the Minkowski spacetime.
We shall discuss the action \eqref{NLSM_action} as the effective action inside the Vainshtein radius.

%%%%%%%%%%%%%%%%%%%%%%%%%%%%%%%%%%%%%%%%%%%%%%%%%%%%%%%%%
%%%%%%%%%%%%%%%%%%%%%%%%%%%%%%%%%%%%%%%%%%%%%%%%%%%%%%%%%
%%%%%%%%%%%%%%%%%%%%%%%%%%%%%%%%%%%%%%%%%%%%%%%%%%%%%%%%%
%%%%%%%%%%%%%%%%%%%%%%%%%%%%%%%%%%%%%%%%%%%%%%%%%%%%%%%%%
%%%%%%%%%%%%%%%%%%%%%%%%%%%%%%%%%%%%%%%%%%%%%%%%%%%%%%%%%
%%%%%%%%%%%%%%%%%%%%%%%%%%%%%%%%%%%%%%%%%%%%%%%%%%%%%%%%%
%%%%%%%%%%%%%%%%%%%%%%%%%%%%%%%%%%%%%%%%%%%%%%%%%%%%%%%%%

\section{Scalar mode instability: general background}
\label{sec_scalar_instability}

In this section, we show that the scalar graviton generally suffers from a ghost and/or a gradient instability
when there is no vector graviton excitation in the massive gravity nonlinear sigma model.
To see the existence of the instability,
we assume a weak gravitational field and ignore the vector graviton excitation:
\begin{align}
g^{\rm GR}_{\mu\nu}&=\eta_{\mu\nu}+h^{\rm GR}_{\mu\nu}\,, \\
\pi_{\mu}&=\partial_{\mu} \pi \,.
\end{align}

In a similar way to the $\Lambda_3$ decoupling limit \cite{deRham:2010ik},
the action can be expanded as
\begin{align}
\mathcal{L}_{\rm MG-NLS}
&=
-\frac{\Lambda_2^4}{2}h^{\rm GR\; \mu\nu} 
\left( X^{(1)}_{\mu\nu}+\beta_2 X^{(2)}_{\mu\nu}+\beta_3 X^{(3)}_{\mu\nu} \right) 
\nn
&\quad +\mathcal{O}(h_{\rm GR}^2)\,,
\end{align} 
where 
\begin{align}
\beta_2&=c_3-1
\,,\quad
\beta_3=-(c_3+c_4)
\,,
\end{align}
and
\begin{align}
X^{(1)}_{\mu\nu}&=-\frac{1}{2!}\epsilon_{\mu}{}^{\alpha\beta\gamma} \epsilon_{\nu \alpha'\beta\gamma} \Pi^{\alpha'}_{\alpha}
\nn
&=[\Pi]\eta_{\mu\nu} -\Pi_{\mu\nu}\,, \label{def_X1}
\\
X^{(2)}_{\mu\nu}&=-\frac{1}{2!}\epsilon_{\mu}{}^{\alpha\beta\gamma} \epsilon_{\nu \alpha'\beta'\gamma} \Pi^{\alpha'}_{\alpha} \Pi_{\beta'}^{\beta}
\nn
&=\frac{1}{2}\eta_{\mu\nu} \left( [\Pi]^2-[\Pi^2] \right)+\Pi^2_{\mu\nu}-[\Pi]\Pi_{\mu\nu}\,, \label{def_X2}
\\
X^{(3)}_{\mu\nu}&=-\frac{1}{3!}\epsilon_{\mu}{}^{\alpha\beta\gamma} \epsilon_{\nu \alpha'\beta'\gamma'} \Pi^{\alpha'}_{\alpha}\Pi^{\beta'}_{\beta}\Pi^{\gamma'}_{\gamma}
\nn
&=\frac{1}{6}\left( [\Pi]^3-3[\Pi][\Pi^2]+2[\Pi^3] \right)  \eta_{\mu\nu}
\nn
&\quad
-\Pi^3_{\mu\nu}+[\Pi]\Pi^2_{\mu\nu}-\frac{1}{2}([\Pi]^2-[\Pi^2])
\,. \label{def_X3}
\end{align}
%with 
Here, we have introduced the notation $\Pi_{\mu\nu}=\partial_{\mu} \partial_{\nu} \pi, 
\Pi^n{}^{\mu}{}_{\nu}=\Pi^{\mu}{}_{\alpha_2} \Pi^{\alpha_2}{}_{\alpha_3} \cdots \Pi^{\alpha_n}{}_{\nu}$
and $[\Pi^n]=\Pi^n{}^{\mu}{}_{\mu}$.
Note that, differently from the case of $\Lambda_3$ decoupling limit,
$h^{\rm GR}_{\mu\nu}$ has been already fixed
and it acts as an external force for the field $\pi$.

The field $\pi$ can be split  into the background configuration $\bar{\pi}$ and the perturbation $\delta \pi$ as
\begin{align}
\pi=\bar{\pi}+ \delta \pi\,,
\end{align}
with $ \delta \pi \ll \bar{\pi}$.
%The background configuration 
$\bar{\pi}$ is determined by the
the equation of motion
\begin{align}
&
\epsilon^{\alpha\beta\gamma\delta} \epsilon_{\mu \nu \rho \sigma}R^{(1)\mu\nu}{}_{\alpha\beta}
\nn
&\times
\left( \delta^{\rho}{}_{\gamma} \delta^{\sigma}{}_{\delta}
+2\beta_2 \bar{\Pi}^{\rho}{}_{\gamma} \delta^{\sigma}{}_{\delta}
+\beta_3 \bar{\Pi}^{\rho}{}_{\gamma} \bar{\Pi}^{\sigma}{}_{\delta} \right)\nn
=\:&\epsilon^{\alpha\beta\gamma\delta} \epsilon_{\mu \nu \rho \sigma} \partial^{\mu}
\Big[ (\partial_{\beta} h^{\rm GR}{}^{\nu}{}_{\alpha}-\partial_{\alpha} h^{\rm GR}{}^{\nu}{}_{\beta})
\nn
&\times
\left( \delta^{\rho}{}_{\gamma} \delta^{\sigma}{}_{\delta}
+2\beta_2 \bar{\Pi}^{\rho}{}_{\gamma} \delta^{\sigma}{}_{\delta}
+\beta_3 \bar{\Pi}^{\rho}{}_{\gamma} \bar{\Pi}^{\sigma}{}_{\delta}
\right) \Big]
=0\,,
\label{background_pi_eom}
\end{align}
where
\begin{align}
R^{(1)}_{\mu\nu\alpha\beta}=
\partial_{\mu}\partial_{[\beta}h_{\alpha]\nu}^{\rm GR}+\partial_{\nu}\partial_{[\alpha}h_{\beta]\mu}^{\rm GR}\,,
\end{align}
is the linearized Riemann curvature and $\bar{\Pi}_{\mu\nu}=\partial_{\mu}\partial_{\nu}\bar{\pi}$.
Then the quadratic order action for the perturbation $\delta \pi$ is given by
\begin{align}
\mathcal{L}_2=-\frac{1}{2}Z^{\mu\nu} \partial_{\mu} \delta \pi \partial_{\nu}  \delta \pi
+\mathcal{O}(h_{\rm GR}^2)\,,
\label{phi_action}
\end{align}
where
\begin{align}
Z^{\mu\nu}&=
-\frac{\Lambda_2^4}{4}\epsilon^{\mu \alpha\beta\gamma}\epsilon^{\nu}{}_{\alpha'\beta'\gamma'}
R^{(1)\alpha'\beta'}{}_{\alpha\beta} 
\nn
&\qquad \qquad \times ( \beta_2 \delta^{\gamma'}{}_{\gamma}+\beta_3 \bar{\Pi}^{\gamma'}{}_{\gamma} )
\,.
\end{align}
We note 
\begin{align*}
Z^{\mu\nu}&=\beta_2 \times ({\rm Ricci\; curvatures})
\nn
&\quad +\beta_3 \times\bar{\Pi} \times  ({\rm Riemann\; and\; Ricci\; curvatures}),
\end{align*}
thus $Z^{\mu\nu}$ is identically zero for a Ricci flat spacetime when $\beta_3=0$.
For this case we should take into account next order contributions of $h_{\rm GR}$.
In this paper, however, we restrict our analysis to the case of $\beta_3 \neq 0$
and assume $Z^{\mu\nu}$ is not zero.

The no-ghost and no-gradient instability condition
is given by the signs of eigenvalues of $Z^{\mu\nu}$
are $[-,+,+,+]$, which
is equivalent to all eigenvalues of $Z^{\mu}{}_{\nu}$ are positive. 
Hence we obtain
\begin{align}
Z^{\mu}{}_{\mu}>0\,,
\end{align}
as a necessary condition of no-instabilities.
However we obtain
\begin{align*}
Z^{\mu}{}_{\mu}\propto {\rm Ricci\; curvatures}\,.
\end{align*}
Since
the sum of the eigenvalues is zero for any Ricci flat spacetime,
there is at least one negative eigenvalue of $Z^{\mu}{}_{\nu}$,
which leads a ghost instability or a gradient instability.
As a result a ghost and/or a gradient instability appears for any Ricci flat background.

For instance, the static and spherically symmetric solution is given by
\begin{align}
h^{\rm GR}_{tt}&=\frac{2GM}{r}\,, \;h^{\rm GR}_{rr}=\frac{2GM}{r}\,, \;{\rm others}=0\,, \\
\partial_{\mu} \bar{\pi}&=(0,r \mu(r),0,0)\,,\:
\mu=\pm \frac{1}{\sqrt{\beta_3}}+\mathcal{O}(GM/r)\,.
\end{align}
Then the matrix $Z^{\mu\nu}$ is given by
\begin{align}
Z^{\mu\nu}=\frac{GM\sqrt{\beta_3}}{r^3} \times {\rm diag}\left[0, \mp 2 ,\pm \frac{1}{r^2} ,\pm \frac{1}{r^2\sin ^2 \theta} \right].
\end{align}
Hence the gradient instability appears from either the radial 
or the angular derivatives. Note that, since the $(tt)$-component of $Z^{\mu\nu}$ is zero at leading order of $GM/r$, it seems that the scalar graviton is infinitely strong coupled. However the kinetic term indeed appears at the next order of $GM/r$ and our effective action does not lose any degrees of freedom as we will see in the next section.

Since the bigravity theory contains degrees of freedom of the vector graviton as well as one of the scalar graviton
and these are coupled to each other in a general background,
one cannot directly conclude the Vainshtein screening solution is unstable in a Ricci flat background spacetime.
Therefore we shall discuss general perturbations including vector gravitons around the static and spherically symmetric background in the next section.
Regardless of this, our result suggests that the Vainshtein screening solutions cannot be supported only by the scalar graviton
and the excitation of the vector graviton has to be taken into account.

Note that our result can be also straightforwardly applied to 
the Horndeski theories, which was discussed in
\cite{Koyama:2013paa}.
Based on the effective action for the Vainshtein mechanism, the paper \cite{Koyama:2013paa} showed
the static and spherically symmetric solution with the Vainshtein screening is unstable 
as long as the Horndeski theory includes so-called $\mathcal{L}_5$ term.
Actually, for the case of Horndeski theory, although other terms appear in the effective action,
they are shown to be sub-dominant if we assume that the Vainshtein screening is working, which gives the  same action as \eqref{phi_action}
(see Appendix \ref{appendix}).

%%%%%%%%%%%%%%%%%%%%%%%%%%%%%%%%%%%%%%%%%%%%%%%%%%%%%%%%%
%%%%%%%%%%%%%%%%%%%%%%%%%%%%%%%%%%%%%%%%%%%%%%%%%%%%%%%%%
%%%%%%%%%%%%%%%%%%%%%%%%%%%%%%%%%%%%%%%%%%%%%%%%%%%%%%%%%
%%%%%%%%%%%%%%%%%%%%%%%%%%%%%%%%%%%%%%%%%%%%%%%%%%%%%%%%%
%%%%%%%%%%%%%%%%%%%%%%%%%%%%%%%%%%%%%%%%%%%%%%%%%%%%%%%%%
%%%%%%%%%%%%%%%%%%%%%%%%%%%%%%%%%%%%%%%%%%%%%%%%%%%%%%%%%
%%%%%%%%%%%%%%%%%%%%%%%%%%%%%%%%%%%%%%%%%%%%%%%%%%%%%%%%%

\section{Instability of static and spherically symmetric solution}
\label{sec_instability_SSS}
In the previous section, we have shown that the Vainshtein screening solutions cannot
be supported only by the excitation of the scalar graviton.
However since the massive gravity nonlinear sigma model contains the degrees of freedom of the vector graviton as well, one cannot still conclude the solution is indeed unstable. In this section, thus, we study the general perturbations around the static and spherically symmetric solutions with Vainshtein screening given by \cite{Babichev:2013pfa,Enander:2015kda,Aoki:2016eov}  in which there exists only the scalar graviton in the background solution.
For simplicity, we only focus on outside the source ($r>R_{\star}>2GM$)
thus the $g$-spacetime is given by Schwarzschild spacetime.
We choose the spherical coordinate
\begin{align}
g^{\rm GR}_{\mu\nu}&={\rm diag} [-F_g^2,F_g^{-2},r^2,r^2\sin^2 \theta]\,,
\label{g_Sch}
\\
\eta_{\mu\nu}&={\rm diag} [-1,1,r^2,r^2\sin^2 \theta]\,,
\label{f_Min}
\end{align}
with $F_g^2=1-\frac{2GM}{r}$ where $M$ and $R_{\star}$ is a gravitational mass and a radius of the star, respectively.

First we give the static and spherically symmetric solution.
The solution can be found by assuming the St\"ueckelberg field as
\begin{align}
\pi_{\mu}=\bar{\pi}_{\mu}=(0,r \mu(r),0,0)\,.
\end{align}
The basic equation can be derived by varying the action with respect to $\mu$.
Then the solution is given by 
\begin{widetext}
\begin{align}
\mu&=\frac{-(1-F_g)(\beta_2-2\beta_3)+\epsilon \sqrt{
(\beta_2-2\beta_3)^2(1-F_g)^2+\beta_3(1+F_g)(4\beta_2-1+(3-4\beta_2)F_g)}
 }{\beta_3(1+F_g)} \nn
&=\frac{\epsilon}{\sqrt{\beta_3}}+\mathcal{O}(GM/r)\,,
\label{mu_sol}
\end{align}
\end{widetext}
where $\epsilon=\pm 1$.
Note that although $\beta_2$ and $\beta_3$ are originally free parameters, it was shown that this solution exists only for $\beta_3 > 1$ 
with some other constraints on $\beta_2$ and $\beta_3$ in bigravity
\cite{Enander:2015kda,Aoki:2016eov}.
Note also that, although the solution \eqref{mu_sol} with the metrics \eqref{g_Sch} and \eqref{f_Min} is obtained by the effective action \eqref{NLSM_action},
this solution is indeed the approximate solution deep inside the Vainshtein radius in the bigravity theory \cite{Aoki:2016eov}.
Here the minus branch $(\epsilon=-1)$ is the asymptotically flat branch.
On the other hand the plus branch $(\epsilon=1)$ is not regular in general, however
the plus branch can describe a regular asymptotically AdS solution 
when we introduce a negative cosmological constant.

We shall study the stability of this solution.
Since the background spacetime is spherically symmetric,
perturbations can be decomposed into odd parity perturbations and
even parity perturbations, that is, 
\begin{align}
\pi_\mu = \bar{\pi}_\mu + \delta \pi^{\rm odd} _\mu +  \delta \pi^{\rm even} _\mu.
\end{align}

These perturbations are decoupled at the linear order equation of motion
(or equivalently at the quadratic order Lagrangian). 
Hence we separately discuss the odd and even parity perturbations, in order.

\subsection{Odd parity perturbations}
First we discuss the odd parity perturbations.
Because of the spherical symmetry of the background solution,
we can separate the variables
%the separation of variables can be used,
and then $x^i=(\theta,\varphi)$ dependence can be expanded in terms of
the vector spherical harmonics $Y_i$, which is defined by
\begin{align}
[D^2+\ell(\ell+1)-1]Y_i&=0\,,\;(\ell=1,2,\cdots)\,,\\
D^iY_i&=0\,.
\end{align}
Here, $D_i$ is the covariant derivative on the 2-sphere
and $D^2=D_iD^i$.
The explicit form of $Y_i$ is given by
\begin{align}
Y_i=\epsilon_{ij}D^jY
\,,
\end{align}
where $\epsilon_{ij}$ is the Levi-Civita tensor and $Y$ is the spherical harmonics satisfying 
\begin{align}
[D^2+\ell(\ell +1)]Y=0\,,\;
(\ell=0,1,2,\cdots)\,.
\end{align}

By using the vector harmonics, 
the perturbation of the St\"ueckelberg field is
expressed by 
\begin{align}
\delta \pi ^{\rm odd}_{\mu}=(0,0,r\chi_{\Omega} Y_i)
\,,
\end{align}
where $\chi_{\Omega}$ is 
 a function of $(t,r)$.

The quadratic order action is given by
\begin{widetext}
\begin{align}
S^{\rm odd}&=
\int r^2dtdrd\Omega
\frac{\Lambda^4_2}{4}(\sqrt{\beta_3}+\epsilon \beta_2)
\left[
\frac{4}{2\sqrt{\beta_3}+\epsilon} \dot{\chi}_{\Omega}^2
-\frac{1}{\sqrt{\beta_3}+\epsilon}
\left(\chi_{\Omega}'{}^2+
\frac{\ell(\ell+1)}{r^2}\chi_{\Omega}^2
\right)
\right]Y_iY^i
+\mathcal{O}(GM/r)
\,,
\end{align}
\end{widetext}
where a dot and a prime denote the time derivative and the 
radial derivative, respectively. Since each eigenmode of the harmonics does not couple with the other eigenmodes, we drop the summation sign.
The stability condition (no-ghost and no-gradient instability) is given by
\begin{align}
\sqrt{\beta_3}+\beta_2>0\,.
\end{align}
for the plus branch, 
while for the minus branch 
the condition is
\begin{align}
\sqrt{\beta_3}-\beta_2>0\,.
\end{align}

\subsection{Even parity perturbations}
Next we consider the even parity perturbations.
By using the spherical harmonics, the perturbation of the St\"ueckelberg field is
expressed by 
\begin{align}
\delta \pi ^{\rm even}_{\mu}&=(\xi_t Y,\xi_r Y,r\xi_{\Omega} D_i Y)\,,
\end{align}
where $\xi_t$, $\xi_r$ and $\xi_{\Omega}$ are functions of $(t,r)$. 
Note that for the $\ell=0$ mode,
the variable $\xi_{\Omega}$ is undefined because $D_iY=0$.
Hence we should discuss $\ell=0$ mode and $\ell\geq 1$ modes, separately.

\subsubsection{Radial perturbation $(\ell=0)$}
For $\ell=0$, the spherical harmonics is simply given by $Y|_{\ell=0}=1/\sqrt{4\pi}$.
The quadratic order action can be schematically expressed by
\begin{align}
S^{\ell=0}=S^{\ell= 0}
(\dot{\xi}_r,\xi_t',\xi_r',\xi_t,\xi_r)
\,,
\end{align}
from which
$\xi_t$ is non-dynamical and it can be integrated out.
The variation with respect to $\xi_t$ yields a constraint equation
\begin{widetext}
\begin{align}
&\partial_t
\left[
2rF_g^{-1}(F_g-1)(\beta_2+\beta_3 \mu) \xi_r \right]
-\partial_r
\left[ \frac{r^2 F_g (1+2\beta_2 \mu+\beta_3 \mu^2)(F_g^2 \xi_t'-\dot{\xi}_r)}{1+F_g^2(r+r\mu)'}
\right]=0\,,
\end{align}
where the solution is given by
\begin{align}
\xi_t'&=-2GM(\beta_2+\epsilon \sqrt{\beta_3}) \Biggl[ \frac{\dot{ { \Xi'}}}{2rF_g(F_g-1)(\beta_2+\beta_3\mu)}
+\frac{1+F_g^2(r+r\mu)'}{r^2F_g^3(1+2\beta_2\mu+\beta_3\mu^2)} \dot{ \Xi }
\Biggl]
\,,
\\
\xi_r&=-2GM(\beta_2+\epsilon \sqrt{\beta_3})  \frac{F_g}{2r(F_g-1)(\beta_2+\beta_3\mu)}   \Xi'
\,,
\end{align}
\end{widetext}
with some function  $\Xi (t,r)$. Here  the factor is introduced so that $\xi_t$ and $\xi_r$ can be expressed by
\begin{align}
\xi_t&=\dot{ \Xi}
+\mathcal{O}(GM/r)
\,,\\
\xi_r
&= \Xi'
+\mathcal{O}(GM/r)
\,,
\end{align}
at  the leading order of $GM/r$.
Then the quadratic action is expressed by
\begin{align}
S^{\ell=0}&=\Lambda_2^4\int r^2dtdr \left[K_t\dot{ \Xi}^2-K_r \Xi'{}^2 \right]
\,,
\end{align}
where
\begin{align}
K_t&=-\epsilon \left(\frac{GM}{r}\right)^2\frac{3\sqrt{\beta_3}(\beta_2+\epsilon \sqrt{\beta_3})}{r^2}
\nn
&\quad +\mathcal{O}\left(\left(\frac{GM}{r}\right)^3\right)
\\
K_r&=-\epsilon \frac{GM}{r}\frac{\sqrt{\beta_3}}{r^2}
+\mathcal{O}\left(\left(\frac{GM}{r}\right)^2\right)
\,.
\end{align}
Note that, while the gradient term appears at the first order of $GM/r$ (i.e., the first order of the metric perturbation around the Minkowski spacetime), the kinetic term appears at the second order of $GM/r$. Hence the scalar graviton is not infinitely strong coupled although the propagation speed is superluminal.

From the second order action, we can see that the plus branch suffers from the gradient instability.
Even for the minus branch, the stability condition is given by
\begin{align}
\beta_2-\sqrt{\beta_3}>0
\,,
\end{align}
which 
has a sign opposite to 
the stability condition of the odd parity perturbations.
As a result we conclude that the static spherically symmetric solution is unstable for any parameters of $\beta_2$ and $\beta_3$.

\subsubsection{General modes $(\ell\geq 1)$}
\label{ell>1}
Although we have shown the instability of the background solution,
we discuss general modes of the even parity perturbations for completeness.
The quadratic action can be expressed by
\begin{align}
S^{\rm even}=\int Y^2 d\Omega \int dtdr \mathcal{L}^{\rm even}
(\dot{\xi}_r,\dot{\xi}_{\Omega},\xi_A',\xi_A)
\,,
\end{align}
where $A=(t,r,\Omega)$,
thus $\xi_t$ is a non-dynamical variable, same as the case of $\ell=0$ mode.
However, 
contrary to  the case of $\ell=0$ mode, 
the constraint equation, which is derived by the variation with respect to $\xi_t$,
is not easily solved.
We notice however that the constraint equation has a particular solution
\begin{align}
 \delta \pi^{\rm even}_{\mu}=\partial_{\mu} (\Xi(t,r) Y(\theta,\varphi))+\mathcal{O}(GM/r)\,,
\end{align}
in which there is no degree of freedom of the vector graviton. Since the stability of the case of the purely scalar graviton has 
already been discussed in the previous section, we shall not discuss this case furthermore here.

To discuss the stability of the general perturbations,
we use the Hamiltonian formulation
and calculate the on-shell Hamiltonian.
The canonical momenta are defined by
\begin{align}
\pi_A&=\frac{\delta \mathcal{L}^{\rm even}}{\delta \dot{\xi}_A}
\,,
\end{align}
Since the Lagrangian does not contain $\dot{\xi}_t$,
there is a primary constraint
\begin{align}
\Phi^1:=\pi_t\approx 0\,,
\end{align}
where the symbol ``$\approx$'' is the weak equality which holds on shell.
$\dot{\xi}_r$ and $\dot{\xi}_{\Omega}$ can be expressed in terms of canonical variables.
Then the total Hamiltonian is given by
\begin{align}
\mathcal{H}_T^{\rm even}
&=\pi_r \dot{\xi}_r+\pi_{\Omega}\dot{\xi}_{\Omega}-\mathcal{L}^{\rm even}
+\lambda \pi_t
\nn
&=\mathcal{H}^{\rm even}[\pi_r,\pi_{\Omega},\xi_A]
+\lambda \pi_t
\end{align}
where $\lambda$ is the Lagrangian multiplier.
The preservation of the primary constraint yields
\begin{align}
\Phi^2:=\{ \Phi^1,H_T^{\rm even} \}\approx 0
\,,
\end{align}
where 
\begin{align}
H_T^{\rm even}=\int dr \mathcal{H}_T^{\rm even}
\,.
\end{align}
Note that since $\Phi^2$ contains only $\xi_t,\pi_r,\pi_r{}',\pi_{\Omega}$, 
the secondary constraint $\Phi^2 \approx 0$ is the constraint equation on the canonical variables,
from which we can easily express  $\xi_t$  in terms of $\pi_r,\pi_r',\pi_{\Omega}$.
This system has only these two constraints.
Indeed, the condition $\{\Phi^2,H_T^{\rm even}\}\approx 0$ contains the Lagrangian multiplier $\lambda$
and
it does not generate a constraint equation,
but a equation  to determine the Lagrangian multiplier.
As a result, we have two constraint equations on the canonical variables which are second class.
Hence the degree of freedom of this system  in the phase space is
\begin{align*}
{\rm d.o.f.}
=6-2=2\times 2
\,,
\end{align*}
which indicates that 
the even parity perturbations contain 
one scalar graviton and one vector graviton.

\begin{widetext}
Substituting the solutions of constraint equations into the Hamiltonian,
the on-shell Hamiltonian is given by
\begin{align}
\quad H_{\rm on-shell}^{\rm even}
&=\int dr \mathcal{H}_{\rm on-shell}^{\rm even}
(\pi_r,\pi_r',\pi_{\Omega},\xi_r,\xi_{\Omega},\xi_{\Omega}')
\,,
\nn
&=\int dr \Lambda_2^4\Big[
\frac{K_1}{r^2} (\pi_r+A_1 \pi_{\Omega})^2
+
\frac{K_2}{r^2} (r\pi_r'+A_2 \pi_{\Omega})^2
+
\frac{K_3}{r^2} \pi_{\Omega}^2
\nn
&\qquad\qquad 
+K_4(\xi_{\Omega}+A_4 \xi_r)^2
+K_5(r\xi_{\Omega}'+A_5 \xi_r)^2
+K_6\xi_r^2
\Big]
\,,
\end{align} 
where
the dimensionless coefficients are expanded as
\begin{align}
K_1&=\epsilon \frac{\mathcal{B}_1}{48\beta_3^{3/2}(\beta_2+\epsilon \sqrt{\beta_3})}
+\mathcal{O}\left(\frac{GM}{r}\right)
\,,\\
K_2&=-\epsilon \left(\frac{GM}{r}\right)^{-2}
\frac{1}{12\sqrt{\beta_3}(\beta_2+\epsilon \sqrt{\beta_3})}
+\mathcal{O}\left( \left(\frac{GM}{r}\right)^{-1} \right)
\,,\\
K_3&=-\epsilon \left(\frac{GM}{r}\right)^{-2}
\frac{\sqrt{\beta_3}(1-2\epsilon \sqrt{\beta_3})^2}{3\mathcal{B}_1(\beta_2+\epsilon \sqrt{\beta_3})}
+\mathcal{O}\left( \left(\frac{GM}{r}\right)^{-1} \right)
\,,\\
K_4&=
-\epsilon \ell(\ell+1)  \frac{GM}{r}
\frac{\mathcal{B}_2}{16\sqrt{\beta_3}(\epsilon+\sqrt{\beta_3})^2}
+\mathcal{O}\left( \left(\frac{GM}{r}\right)^2 \right)
\,,\\
K_5&=
\epsilon \ell(\ell+1) \frac{\beta_2+\epsilon \sqrt{\beta_3}}{4(\epsilon+\sqrt{\beta_3})}
+\mathcal{O}\left(\frac{GM}{r}\right)
\,,\\
K_6&=\epsilon \ell(\ell+1) \left(\frac{GM}{r}\right)^{-1} \sqrt{\beta_3} \mathcal{B}_2^{-1} (\beta_2+\epsilon \sqrt{\beta_3})^2
+\mathcal{O}(1)
\,,
\end{align}
and
\begin{align}
A_1&=\left(\frac{GM}{r}\right)^{-1}4\beta_3 \mathcal{B}_1^{-1}(1-2\epsilon \sqrt{\beta_3})
+\mathcal{O}(1)
\,,\\
A_2&=-1+\mathcal{O}\left(\frac{GM}{r}\right)
\,,\\
A_4&=\left(\frac{GM}{r}\right)^{-1}4\sqrt{\beta_g}\mathcal{B}_2^{-1}(\epsilon+\sqrt{\beta_3})(\beta_2+\epsilon \sqrt{\beta_3})
+\mathcal{O}(1)
\,,
\\
A_5&=-1+\mathcal{O}\left(\frac{GM}{r}\right)
\end{align}
with 
\begin{align}
\mathcal{B}_1&:=\beta_2+8\beta_3-4\beta_2\beta_3+\epsilon\sqrt{\beta_3}(4\beta_3-3)
\,,\\
\mathcal{B}_2&:=\beta_2^2(1-4\beta_3)+\beta_3(5+4\beta_3)-4\beta_2\beta_4
+2\epsilon \sqrt{\beta_3}(6\beta_3-\beta_2(1+4\beta_3))
\end{align}
One can find $K_1K_3<0$ and $K_4K_6<0$ for any parameters $(\beta_2,\beta_3)$,
thus the Hamiltonian is unbounded from the below, 
which means that the perturbations suffer from the instability.
\end{widetext}

%%%%%%%%%%%%%%%%%%%%%%%%%%%%%%%%%%%%%%%%%%%%%%%%%%%
%%%%%%%%%%%%%%%%%%%%%%%%%%%%%%%%%%%%%%%%%%%%%%%%%%%
%%%%%%%%%%%%%%%%%%%%%%%%%%%%%%%%%%%%%%%%%%%%%%%%%%%
%%%%%%%%%%%%%%%%%%%%%%%%%%%%%%%%%%%%%%%%%%%%%%%%%%%
%%%%%%%%%%%%%%%%%%%%%%%%%%%%%%%%%%%%%%%%%%%%%%%%%%%
%%%%%%%%%%%%%%%%%%%%%%%%%%%%%%%%%%%%%%%%%%%%%%%%%%%
%%%%%%%%%%%%%%%%%%%%%%%%%%%%%%%%%%%%%%%%%%%%%%%%%%%

\section{Summary and discussion}
\label{summary}
In this paper, we showed that the massive gravity nonlinear sigma model gives an effective theory of  the vector and scalar gravitons
inside the Vainshtein radius for general massive/bi-gravity. 
We obtained the effective action by taking the $\Lambda_2$ decoupling limit around a curved spacetime and
it can be used as long as we have the Vainshtein screening solutions. 
Making use of the massive gravity nonlinear sigma model as the effective action inside the Vainshtein radius, we studied the stability of the
Vainshtein screening solutions in massive/bi-gravity.

First we  derived
a general consequence that
in any Ricci flat 
background spacetime, the scalar graviton generally suffers from a ghost and/or a gradient instability 
as long as the vector graviton is not excited. Since the spacetime is given by a solution in GR, the Ricci flat region is realized by the vacuum region of the spacetime, thus the instability is found outside 
 the source. 
However since the massive/bi-gravity contains the vector graviton and the perturbations of scalar and vector gravitons are coupled, one cannot directly conclude that 
the Ricci flat Vainshtein screening background spacetime is indeed unstable. 

Hence we 
studied perturbations around the static and spherically symmetric solution obtained in Ref.~\cite{Aoki:2016eov}  next.  
We clarified the stability condition for both odd parity perturbations and even parity perturbations, which depends on $\beta_2$ and $\beta_3$, model parameters
in the massive gravity nonlinear sigma model as well as $\epsilon$, a parameter depending on the asymptotic behavior of the background solution. 
As a result, for any parameters ($\beta_2, \beta_3, \epsilon$), we found that the perturbations suffer from some of the instabilities and
confirmed that the Vainshtein screening background solution is unstable.

We have shown the (local) instability of the spherically symmetric solution in the space region outside the star. In addition, the instability of a black hole solution was shown in \cite{Babichev:2013una,Brito:2013wya} (see also \cite{Kodama:2013rea,Babichev:2014oua,Babichev:2015zub,Babichev:2015xha}). 
Note that our background solution completely differs from the background solution discussed in \cite{Babichev:2013una,Brito:2013wya}. For the black hole solution, both metrics are given by same Schwarzschild metric (or Kerr metric) in which the St\"ueckelberg fields are not excited, i.e., $\phi^a=x^a$. One may expect that there exists a stable hairy black hole supported by hair of the St\"ueckelberg fields. However, our result suggests that a scalar graviton hair is not helpful for supporting astrophysical objects. In particular, existence of a spherically symmetric hairy black hole is unlikely as numerically shown in \cite{Brito:2013xaa}.

The instability implies a difficulty to construct viable astrophysical objects in the context of massive/bi-gravity. 
The universality of the instability suggests that the Vainshtein screening
could not be realized only by the scalar graviton.
%in massive/bi-gravity. 
To obtain a stable solution with the Vainshtein  screening,
the vector graviton has to be nonlinearly excited in a vacuum region of the spacetime. 
Therefore it is also important to study the property of the vector graviton 
in more general spacetimes
for the Vainshtein mechanism in massive/bi-gravity.

%%%%%%%%%%%%%%%%%%%%%%%%%%%%%%%%%%%%%%%%%%%%%%%%%%%%%
%%%%%%%%%%%%%%%%%%%%%%%%%%%%%%%%%%%%%%%%%%%%%%%%%%%%%
%%%%%%%%%%%%%%%%%%%%%%%%%%%%%%%%%%%%%%%%%%%%%%%%%%%%%
%%%%%%%%%%%%%%%%%%%%%%%%%%%%%%%%%%%%%%%%%%%%%%%%%%%%%
%%%%%%%%%%%%%%%%%%%%%%%%%%%%%%%%%%%%%%%%%%%%%%%%%%%%%
%%%%%%%%%%%%%%%%%%%%%%%%%%%%%%%%%%%%%%%%%%%%%%%%%%%%%

%%%%%%%%%%%%%%%%%%%%%%%%%%%%%%%%%%%%%%%%%%%%%%%%%%%%%%%%%%%%%%%
%%%%%%%%%%%%%%%%%%%%%%%%%%%%%%%%%%%%%%%%%%%%%%%%%%%%%%%%%%%%%%%
\section*{Acknowledgments}
%%%%%%%%%%%%%%%%%%%%%%%%%%%%%%%%%%%%%%%%%%%%%%%%%%%%%%%%%%%%%%%
%%%%%%%%%%%%%%%%%%%%%%%%%%%%%%%%%%%%%%%%%%%%%%%%%%%%%%%%%%%%%%%
K.A. would like to thank Shinji Mukohyama for useful discussions and comments.  His work was supported in part by Grants-in-Aid from the Scientific Research Fund of the Japan Society for the Promotion of Science  (No. 15J05540). This work was supported in part by JSPS, Grant-in-Aid for Scientific Research No. 16K17709 (S.M.).

%%%%%%%%%%%%%%%%%%%%%%%%%%%%%%%%%%%%%%%%%%%%%%%%%%
\appendix
\section{Scalar mode instability in the Horndeski theory}
\label{appendix}
In this appendix we briefly discuss the stability of the Vainshtein mechanism in the Horndeski theory, the most general scalar-tensor theory
in the sense that its equation of motion contains up to the second derivative of the fields.
The effective theory for the Vainshtein mechanism around the Minkowski spacetime \cite{Koyama:2013paa} is given by the Lagrangian
\begin{align}
\mathcal{L}=&-\frac{1}{4}h^{\mu\nu}\mathcal{E}_{\mu\nu,\alpha\beta}h^{\alpha\beta}
+\sum_{n=2}^{5}\frac{\tilde{\alpha}_n}{\Lambda_3^{3(n-2)} }\mathcal{L}^{\rm gal}_n
\nn
&-\frac{1}{2}\sum_{n=1}^3 \frac{\tilde{\beta}_n}{\Lambda_3^{3(n-1)}} h^{\mu\nu}X^{(n)}_{\mu\nu}+\frac{1}{2M_{\rm pl}}h_{\mu\nu}T^{\mu\nu} \,,
\end{align}
where
\begin{align}
\mathcal{L}_2^{\rm gal}&=-\frac{1}{2}(\partial \pi)^2\,,\\
\mathcal{L}_3^{\rm gal}&=-\frac{1}{2}(\partial \pi)^2 [\Pi]\,, \\
\mathcal{L}_4^{\rm gal}&=-\frac{1}{2}(\partial \pi)^2 ([\Pi]^2-[\Pi^2 ]) \,, \\
\mathcal{L}_5^{\rm gal}&=-\frac{1}{12}(\partial \pi)^2 ([\Pi]^3-3[\Pi][\Pi^2]+2[\Pi^3])\,,
\end{align}
and $X^{(n)}_{\mu\nu}$ are defined by Eqs \eqref{def_X1}-\eqref{def_X3}.
The dimensionless coefficients $\tilde{\alpha}_n$ and $\tilde{\beta}_n$ are determined by the Horndeski action which are assumed to be order unity.
$\Lambda_3$ is a parameter with the mass dimension which decides the strong coupling scale.
Note that, only in this appendix, the scalar field $\pi$ and the metric perturbation $h_{\mu\nu}$ are normalized to be mass dimension one.
 When the $\mathcal{L}_5$ term in Horndeski theory is non-zero, the parameter $\tilde{\beta}_3$ is generally non-zero which we assume here (see \cite{Koyama:2013paa}).

We assume the existence of the Vainshtein screening solution in 
the spacetime region such as $\partial \partial h \gg \Lambda_3^3$ thus the metric perturbation is locally approximated by a solution of GR.  We also 
assume the scalar field is split into some nonlinear expectation value and a fluctuation: $\pi=\bar{\pi}+\delta \pi$ with $\bar{\pi}\gg \delta \pi$. Then, the linearized Einstein equation is given by
\begin{align}
\mathcal{E}_{\mu\nu,\alpha\beta}h^{\alpha\beta}+
\sum_{n=1}^3 \frac{\tilde{\beta}_n}{\Lambda_3^{3(n-1)}} h^{\mu\nu}X^{(n)}_{\mu\nu}(\bar{\pi})
=\frac{1}{M_{\rm pl}}T_{\mu\nu}\,.
\end{align}
When the Vainshtein mechanism works (i.e., the metric perturbation is approximated by 
the solution of GR), the metric perturbation and the scalar field should satisfy the following inequalities:
\begin{align}
\partial \partial h \gg \partial \partial \bar{\pi} \,,\; \frac{1}{\Lambda_3^3} ( \partial \partial \bar{\pi} )^2\,, \; \frac{1}{\Lambda_3^6} ( \partial \partial \bar{\pi} )^3\,.
\label{Vainshtein_condition}
\end{align}
The quadratic Lagrangian for the scalar fluctuation is given by
\begin{align}
\mathcal{L}_2=-\frac{1}{2}(Z_{\rm gal}^{\mu\nu}+Z_h^{\mu\nu})\partial_{\mu } \delta \pi \partial_{\nu} \delta \pi\,,
\end{align}
where
\begin{align}
Z^{\mu\nu}_{\rm gal}&=\tilde{\alpha}_2\eta^{\mu\nu}
 +\sum_{n=1}^3 \frac{\tilde{\alpha}_{n+2}}{\Lambda_3^{3n}} X^{(n)\mu\nu}(\bar{\pi}) \,,
\\
Z^{\mu\nu}_h&=-\frac{M_{\rm pl}}{4\Lambda_3^3}\epsilon^{\mu \alpha\beta\gamma}\epsilon^{\nu}{}_{\alpha'\beta'\gamma'}
R^{(1)\alpha'\beta'}{}_{\alpha\beta}(h)
\nn
&\qquad \qquad \times \left( \tilde{\beta}_2 \delta^{\gamma'}{}_{\gamma}+\frac{\tilde{\beta}_3}{\Lambda_3^3} \bar{\Pi}^{\gamma'}{}_{\gamma} \right)
\,.
\end{align}
with the linearized Riemann curvature $R^{(1)\alpha'\beta'}{}_{\alpha\beta}(h)\sim \partial \partial h/M_{\rm pl}$. The inequalities \eqref{Vainshtein_condition} and $\partial \partial h \gg \Lambda_3^3$ suggest
\begin{align}
Z^{\mu\nu}_{\rm gal} \ll Z_h^{\mu\nu}\,,
\end{align}
 which means that the discussion in Section \ref{sec_scalar_instability} can be applied to the case of  the Horndeski theory
and then the fluctuation $ \delta \pi$ suffers from a ghost and/or a gradient instability.

\bibliography{ref}

\end{document}